\documentclass[iop]{emulateapj}
\slugcomment{{\sc Accepted for publication in the Astrophysical Journal Letters}}
\usepackage{epsfig, natbib, graphicx}
\usepackage{subfigure}
\def\arcsec{\hbox{$^{\prime\prime}$}}

\begin{document}
\title{\textit{SWIFT} study of the first SGR-like burst from AXP 1E~1841--045 in SNR Kes 73}
\author{Harsha S. Kumar\altaffilmark{1} and Samar Safi-Harb\altaffilmark{1,2}}
\altaffiltext{1}{Department of Physics \& Astronomy, University of Manitoba, Winnipeg, MB R3T 2N2, Canada; harsha@physics.umanitoba.ca}
\altaffiltext{2}{Canada Research Chair; samar@physics.umanitoba.ca}

\begin{abstract}

We report the study of the short (32~ms) and first SGR-like burst observed from the anomalous X-ray pulsar (AXP) 1E 1841--045 associated with the supernova remnant (SNR) Kes 73, discovered on 2010 May 6 by the Burst Alert Telescope (BAT) onboard the \textit{Swift} $\gamma$-ray observatory. The 15--100 keV time-averaged burst spectrum is modeled by a single power-law (PL) with a photon index $\Gamma$=3.2$^{+1.8}_{-1.0}$, and has a fluence of 1.1$^{+0.4}_{-0.6}$$\times$10$^{-8}$ ergs~cm$^{-2}$, luminosity of 2.9$^{+1.1}_{-1.6}$$\times$10$^{39}$ ergs~s$^{-1}$, and energy of 7.2$^{+0.4}_{-0.6}$$\times$10$^{36}$ ergs. The prompt after-burst 0.5--10 keV quiescent spectrum obtained with the \textit{Swift} X-ray Telescope (XRT) is best-fit by an absorbed PL model with  $\Gamma$=2.6$\pm$0.2 and an unabsorbed flux of 9.1$^{+1.2}_{-1.4}$$\times$10$^{-11}$ ergs~cm$^{-2}$~s$^{-1}$. To investigate the pre-burst 0.5--10 keV persistent emission, we analyzed the archival \textit{XMM-Newton} observations and the spectra are well fitted by a two-component blackbody (BB) plus PL model with a temperature $kT$=0.45$\pm$0.03 keV, $\Gamma$=1.9$\pm$0.2, and an unabsorbed flux of 4.3$^{+0.9}_{-1.2}$$\times$10$^{-11}$ ergs~cm$^{-2}$~s$^{-1}$. Comparing the \textit{Swift}-XRT spectrum with the \textit{XMM-Newton} spectrum, spectral softening post-burst is evident with a 2.1 times increase in the unabsorbed flux. We discuss the burst activity and the persistent emission properties of AXP~1E 1841--045 in comparison with other magnetars and in the context of the magnetar model.

\end{abstract}


\section{Introduction}
\label{1}

Recent years have seen many discoveries supporting the idea that anomalous X-ray pulsars (AXPs) and soft $\gamma$-ray repeaters (SGRs) are manifestations of magnetars, the ultra-magnetized ($B$$\sim$10$^{14}$--10$^{15}$~G) isolated neutron stars (NSs) powered by the energy stored in their $B$-field (see Mereghetti 2008 for a recent review). AXPs were identified as sources of persistent X-ray pulsations with periods $P$$\sim$2--12~s, spin-down periods $\dot{P}$$\sim$10$^{-10}$--10$^{-12}$~s~s$^{-1}$, and characteristic spin-down timescales of $\sim$10$^3$--10$^5$ years; whereas SGRs were discovered as sources emitting soft, irregular bursts in the soft $\gamma$-rays.  It was not until a few years ago when the AXPs, known for their X-ray flux variability and glitches, started showing bursting behaviour.  According to the magnetar model (Thompson \& Duncan 1995), AXP outbursts are caused by NS fracturing owing to internal magnetic stresses accompanied by external surface and magnetospheric disturbances. The first SGR-like burst was observed from AXP 1E~1048.1--5937 (Gavriil et al. 2002), followed by 
1E~2259+586, XTE~J1810--197, 4U~0142+61, CXOU~J164710.2--455216 and 1E~1547.0--5408 (Kaspi et al. 2003; Woods et al. 2005; Gavriil et al. 2010; Israel et al. 2007, 2010).  The seventh AXP observed to burst is 1E~1841--045, caught by the Burst Alert Telescope (BAT) onboard the \textit{Swift} $\gamma$-ray observatory (Beardmore et al. 2010).

AXP 1E~1841--045, associated with the young ($\sim$2000~yr) and small ($\sim$4.5$\arcmin$ in diameter) supernova remnant (SNR) Kes 73, has a rotation period $P$=11.8~s, period derivative $\dot{P}$=4.1$\times$10$^{-11}$~s~s$^{-1}$ and a dipole $B$-field$\sim$7.1$\times$10$^{14}$~G (Vasisht \& Gotthelf 1997). A \textit{Chandra} continuous clocking (CC) mode observation of the source resolved the AXP from the surrounding SNR 
and the X-ray spectrum was described by a blackbody (BB, temperature $kT$=0.44$\pm$0.02~keV) plus a power-law (PL, photon index $\Gamma$=2.0$\pm$0.3) model (Morii et al. 2003). The source, long known to be steady, displayed 3 glitches between 1999 and 2008 (Dib et al. 2008) and showed no evidence of glitch-correlated flux changes (Zhu and Kaspi 2010).

In this letter, we report \textit{Swift} observations of the first SGR-like burst detected from AXP 1E~1841--045 with BAT in the 15--100 keV band and the burst-induced changes in the 0.5--10 keV persistent emission with \textit{Swift}'s X-ray Telescope (XRT), together with the two archival \textit{XMM-Newton} observations to investigate the pre-burst persistent emission in the same energy band.  These results are discussed in comparison with other magnetar bursts and in the context of the magnetar model.

\section{Observation and Data Reduction}
\label{2}

\subsection{BAT observations}

\textit{Swift}-BAT (Gehrels et al. 2004; Barthelmy et al. 2005) triggered and located the first burst from 1E~1841--045 (trigger \textit{00421262000}) on 2010 May 06 at 14:37:44 UT (BAT calculated R.A=18h 41m 19s, Dec=$-$04d 55$\arcmin$ 15$\arcsec$ [J2000], with an uncertainty of 3$\arcmin$ radius; Beardmore et al. 2010). The data were analyzed using the standard BAT software distributed within FTOOLS under the HEASoft v6.4.1 package  and the latest calibration files available. The burst pipeline script, \textit{batgrbproduct}, was run to process the BAT trigger event.  In Figure~1, we show the 15--150 keV background-subtracted 4~ms binned light curve of the 32~ms burst 
created using the task \textit{batbinevt}. The burst spectrum and the response matrix were generated using the tasks \textit{batbinevt} and \textit{batdrmgen}, respectively. Finally, a systematic error was applied using the task \textit{batphasyserr} to account for residuals in the response matrix.

\begin{figure}
\includegraphics[width=0.51\textwidth]{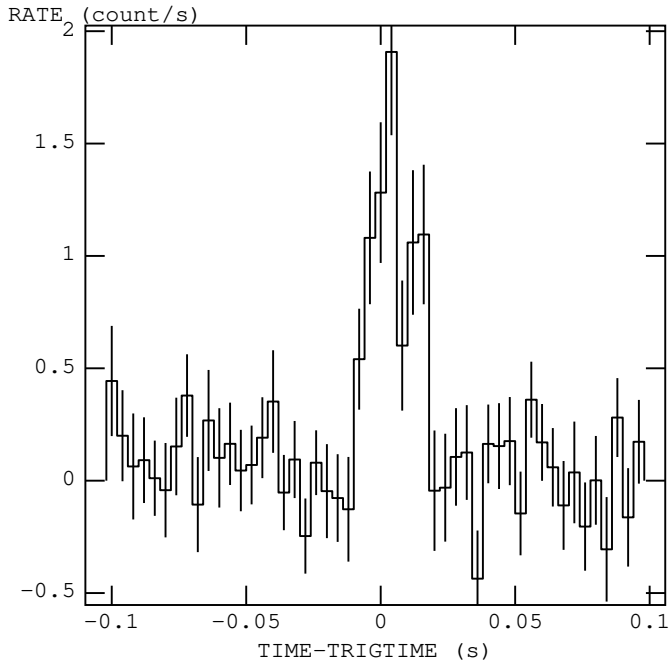}
\caption{The 15--150 keV background-subtracted 4~ms binned light curve of 1E~1841--045's burst detected by \textit{Swift}-BAT.}
\end{figure}

The data were also searched for persistent emission from 1E~1841--045 during the non-bursting intervals. An image was created excluding the burst time intervals and the task \textit{batcelldetect} was run to search for any sources in the BAT sky image. We investigated the time interval $t$=100.51~s to $t$=959.22~s with a net exposure of 859~s. We do not find any significant persistent emission and the 3$\sigma$ upper limit on its 15--100 keV flux is
1.2$\times$10$^{-10}$ ergs~cm$^{-2}$~s$^{-1}$.

\subsection{X-Ray Telescope (XRT) observations}
\label{2.1}

\textit{Swift}-XRT is designed to perform automated observations of newly discovered bursts in the 0.2--10 keV energy band (Burrows et al. 2005). The first XRT observation (\textit{00421262000}) of 1E~1841--045 began at 15:09:28 UT (0.5~hr after the BAT trigger) and the second observation (\textit{00421262002}) started at 16:30:24 UT (1.9~hr after the BAT trigger) on 2010 May 6 for a total exposure of 447~s and 4473~s, respectively.  The XRT data were accumulated in the Windowed Timing (WT) and Photon Counting (PC) mode, however, we considered only the PC mode data which provides full spatial and spectral resolution with a time resolution of 2.5~s. The data were processed using the FTOOLS task \textit{xrtpipeline} v0.12.4. The AXP spectra were extracted from a circular region of radius 20 pixels (1 pixel=2.36$\arcsec$) encompassing 90\% of the encircled energy and the background events were extracted from an annular region of radius between 30 and 50 pixels, centered on the AXP. We used the latest spectral redistribution matrix (RMF; \textit{swxpc0to12s6\_20070901v011.rmf}) available in CALDB. The ancillary response files (ARFs) were generated using the \textit{xrtmkarf} task which accounts for the different extraction regions, vignetting, and point-spread function corrections. 

\subsection{XMM-Newton observations}
\label{2.2}

In order to study the pre-burst quiescent emission, we used the archival \textit{XMM-Newton} observations of 1E~1841--045 made on 2002 October 5 and 7 (ObsIDs: \textit{0013340101}  \& \textit{0013340201}) with the European Photon Imaging Cameras (EPIC) MOS (Turner et al. 2001) operating in full window mode and PN (Struder et al. 2001) operating in larger window mode. We analyzed the data using the $\textit{XMM-Newton}$ Science Analysis System (SAS) v10.0.0 and the most recent calibration files. We created light curves with 100~s bins and the bins with count-rates greater than 0.35 and 0.4 counts~s$^{-1}$ were rejected for MOS1/2 and PN, respectively, thus filtering out spurious and heavy proton flaring events. The total effective exposure times for MOS1+2 and PN cameras were 20.3~ks and 6.7~ks, respectively.

The AXP spectra were extracted from a 20$\arcsec$ circular region from MOS1/2 and PN encompassing 77\% of the encircled energy and the background spectra were extracted from an annular region of radius between 20$\arcsec$ and 30$\arcsec$, centered on the AXP.  These extraction radii were chosen to maximize the emission from the pulsar using all three detectors
while also avoiding contamination from the surrounding bright SNR Kes~73. We found that a larger extraction radius for the pulsar clearly showed emission lines characteristic of a young SNR (see Zhu \& Kaspi 2010 for their analysis of the PN data in the 4--10 keV), thereby introducing uncertainties in accurately determining the spectral properties of the pulsar. Pile up is negligible in both observations. Next, we created RMFs and ARFs for the corresponding detector regions using the commands \textit{rmfgen} and \textit{arfgen}.

Spectral fitting for all data was performed using XSPEC v12.6.0 and the errors quoted are at the 90\% confidence level. The spectra were grouped to have a minimum of 20 and 25 counts per bin for the XRT and \textit{XMM-Newton} data, respectively, in the 0.5--10 keV band. \textit{Swift}-BAT spectral fitting was restricted to the 15--100 keV band since the spectrum above 100 keV was contaminated by noise.

\begin{figure*}[ht]
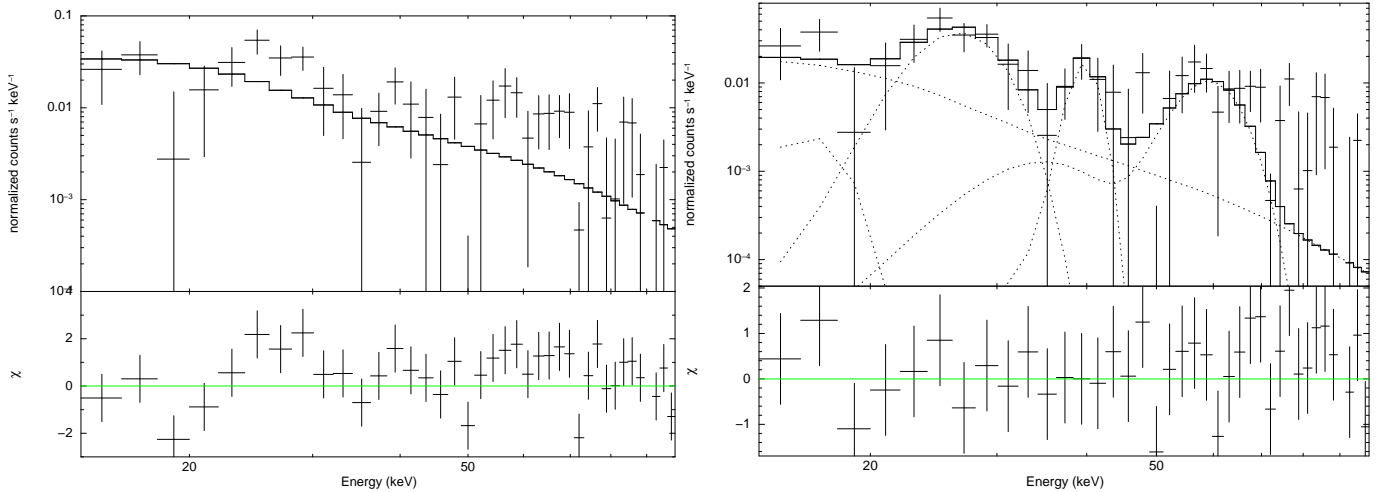

\includegraphics[width=0.36\textwidth, angle=-90]{fig2a.eps}
\includegraphics[width=0.36\textwidth, angle=-90]{fig2b.eps}
\caption{{\it{Left}}: The 15--100 keV spectrum of the 32~ms burst detected with \textit{Swift}-BAT fitted with a PL model ($\chi^2_{\nu}$=1.531, $\nu$=36), showing residuals at energies around 27 keV, 40 keV and 60 keV. {\it{Right}}: Best-fit PL model plus 3 Gaussian lines to account for the emission features in the burst spectrum ($\chi^2_{\nu}$=0.963 (27)).}
\end{figure*}

\section{Analysis and Results}
\label{3}

\subsection{Burst spectroscopy}
\label{3.1}

We fitted the time-averaged burst spectrum with simple models: a PL model with $\Gamma$=2.6$\pm$0.4 (reduced chi-squared $\chi^2_{\nu}$=1.531 (36) where $\nu$ is the number of degrees of freedom) and a BB model with $kT$=8.7$\pm$1.3 keV ($\chi^2_{\nu}$=1.292 (36)).  Neither model gave a good fit, however there were hints of spectral features (Figure~2). Hence, we added Gaussian emission lines to the PL model and found that the addition of three broad 
lines at energies 26.7$^{+1.6}_{-1.4}$ keV, 39.6$^{+2.8}_{-2.2}$ keV and 59.8$^{+3.1}_{-3.5}$ keV improves the fit  ($\chi^2_{\nu}$=0.963 (27)) with an $F$-test probability of 7$\times$10$^{-3}$, and $\Gamma$=3.2$^{+1.8}_{-1.0}$. The best-fit parameters are summarized in Table~1. 
We also obtained an acceptable fit ($\chi^2_{\nu}$=0.928 (27)) by adding Gaussian lines to the BB model with a $kT$=1.7$^{+0.4}_{-0.2}$ keV (BB radius=15.0$^{+2.2}_{-1.9}$~km) and similar line energies (Table~1).  The addition of a second BB-component to the above mentioned models was ruled out statistically owing to higher $\chi^2_{\nu}$ values. At a distance of 8.5 kpc (Tian \& Leahy 2008), we estimate a burst fluence of 1.1$^{+0.4}_{-0.6}$$\times$10$^{-8}$ ergs~cm$^{-2}$, a peak luminosity of 2.9$^{+1.1}_{-1.6}$$\times$10$^{39}$ ergs~s$^{-1}$ and a total energy of 7.2$^{+0.4}_{-0.6}$$\times$10$^{36}$~ergs in the 15--100 keV band. These values are consistent with those obtained for the other AXP bursts observed with \textit{Swift} (Israel et al. 2007, 2010).

In order to further address the statistical significance of the spectral features, we performed a Monte Carlo simulation in XSPEC.
We generated 1000 fake spectra as described in the BAT analysis manual\footnote{http://heasarc.nasa.gov/docs/swift/analysis/threads/batsimspectrumthread.html} using the task \textit{fakeit none} with the same response matrix as our observation and then applied corrections to the simulated spectra using the task \textit{batphasimerr}.   All the simulated spectra were fit with a PL model and a PL plus Gaussian emission lines model searching over the energy ranges 24--28 keV, 37--41 keV, and 57--61 keV in steps of 0.2~keV. For each faked spectrum, we computed the $\chi^2$ difference ($\Delta\chi^2$) between the two models and none of them gave a difference $|\Delta\chi^2|\ge29.11$, the observed value in our fitted data. We conclude that the probability of obtaining the three spectral features by random chance is $<10^{-3}$. Their nature is further discussed in Section~4.

\begin{figure}[h]
\includegraphics[width=0.35\textwidth,angle=-90]{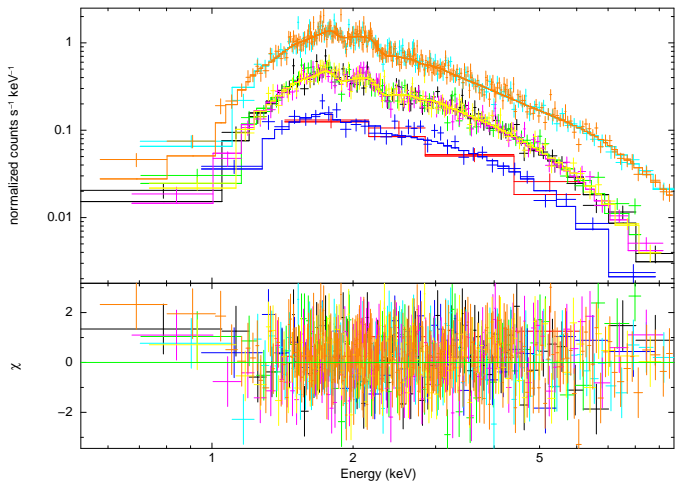}
\caption{Best-fit BB+PL and PL fits to the \textit{XMM-Newton} and \textit{Swift}-XRT data, respectively. The topmost, middle and bottom spectra represent the \textit{XMM-Newton} PN, MOS 1+2 and the two \textit{Swift}-XRT data, respectively.}
\end{figure}

\begin{table*}
\caption{Summary of the best-fit spectral parameters. The X-ray luminosity is at a distance of 8.5~kpc.}
\begin{tabular}{l l l l l l}
\hline\hline
Parameter & \multicolumn{2}{c}{Burst (15--100 keV)} & \multicolumn{3}{c}{Persistent Emission (0.5--10 keV)} \\
\cline{2-6} & & & Pre-burst (\textit{XMM}) & \multicolumn{2}{c}{Post-burst (\textit{XRT})}  \\
\cline{4-6}
& BB & PL & BB+PL  & PL  & BB+PL  \\
\hline
$N_H$ (10$^{22}$ cm$^{-2}$) & \nodata & \nodata & 2.2$^{+0.1}_{-0.1}$ & 2.3$^{+0.3}_{-0.3}$ & 2.2$^{+0.1}_{-0.1}$ \\

$\Gamma$ & \nodata & 3.2$^{+1.8}_{-1.0}$ & 1.9$^{+0.2}_{-0.2}$ & 2.6$^{+0.2}_{-0.2}$ & 2.6$^{+0.2}_{-0.1}$ \\

$kT_{BB}$ (keV) & 1.7$^{+0.4}_{-0.2}$ & \nodata & 0.45$^{+0.03}_{-0.03}$ & \nodata & 0.55$^{+0.23}_{-0.20}$ \\

Line energy, $E_1$ (keV) & 26.5$^{+1.6}_{-1.6}$ & 26.7$^{+1.6}_{-1.4}$ & \nodata & \nodata & \nodata \\

Width, $\sigma_1$ (keV) & 3.1$^{+2.1}_{-1.5}$ & 2.7$^{+1.9}_{-1.3}$ & \nodata & \nodata & \nodata\\

Line energy, $E_2$ (keV) & 39.8$^{+2.6}_{-7.4}$ & 39.6$^{+2.8}_{-2.2}$ & \nodata & \nodata & \nodata \\

Width, $\sigma_2$ (keV) & 1.3$^{+2.7}_{-1.2}$ & 0.6$^{+3.2}_{-0.4}$ & \nodata & \nodata & \nodata  \\

Line energy, $E_3$ (keV) & 59.9$^{+2.6}_{-3.7}$ & 59.8$^{+3.1}_{-3.5}$ & \nodata & \nodata & \nodata \\

Width, $\sigma_3$ (keV) & 5.2$^{+2.5}_{-1.9}$ &  4.9$^{+2.3}_{-1.7}$ & \nodata & \nodata & \nodata \\

$\chi^2_{\nu}$ & 0.928 (27) & 0.963 (27) & 0.931 (1267) & 1.025 (87) & 1.034 (85) \\

\hline
$F_{unabs}$ (ergs~cm$^{-2}$~s$^{-1}$) &  3.4$^{+1.5}_{-1.2}$$\times$10$^{-7}$ & 3.4$^{+1.3}_{-1.9}$$\times$10$^{-7}$ & 4.3$^{+0.9}_{-1.2}$$\times$10$^{-11}$ & 9.1$^{+1.2}_{-1.4}$$\times$10$^{-11}$ & 8.1$^{+3.7}_{-2.7}$$\times$10$^{-11}$\\

$L_x$ (ergs s$^{-1}$) & 2.9$^{+1.3}_{-1.0}$$\times$10$^{39}$ & 2.9$^{+1.1}_{-1.6}$$\times$10$^{39}$ & 3.7$^{+0.8}_{-1.0}$$\times$10$^{35}$ & 7.8$^{+1.0}_{-1.2}$$\times$10$^{35}$ & 7.0$^{+3.2}_{-2.3}$$\times$10$^{35}$ \\

\hline 
\end{tabular}
\end{table*}

\subsection{Persistent post-burst and pre-burst emission}
\label{3.2}

Spectral modeling was performed by fitting together the two \textit{Swift}-XRT observations taken immediately after the BAT trigger, since the \textit{00421262000} data had less counts (132$\pm$13) than the \textit{00421262002} data (1480$\pm$44). We first fitted the data with an
absorbed BB model which did not provide an acceptable fit ($\chi^2_{\nu}$=1.419 (87)) yielding a low $N_H$=(0.8$\pm$0.2)$\times$10$^{22}$ cm$^{-2}$ and $kT$=0.85$\pm$0.05~keV, with the high-energy end of the spectra poorly characterized. Next, we fitted an absorbed PL model which yielded a good fit ($\chi^2_{\nu}$=1.025 (87)) with $N_H$=(2.3$\pm$0.3)$\times$10$^{22}$ cm$^{-2}$ and $\Gamma$=2.6$\pm$0.2.  We also considered the inclusion of a second BB-component (generally required to describe the magnetar persistent spectra) to the PL model which gave an $N_H$=(2.2$\pm$0.1)$\times$10$^{22}$ cm$^{-2}$, $\Gamma$=2.6$^{+0.2}_{-0.1}$, and $kT$=0.55$^{+0.23}_{-0.20}$~keV with  $\chi^2_{\nu}$=1.034 (85). This fit is also acceptable, but an $F$-test probability of 0.62413 suggests that the second BB-component is not statistically needed.

We also analyzed the archival \textit{XMM-Newton} observations to investigate the pre-burst spectrum.  A PL+BB model yields a good fit ($\chi^2_{\nu}$=0.931 (1267)) with the following parameters:  $N_H$=(2.2$\pm$0.1)$\times$10$^{22}$ cm$^{-2}$, $\Gamma$=1.9$\pm$0.2, and $kT$=0.45$\pm$0.03~keV (inferred radius $R_{BB}$=5.0$^{+0.5}_{-0.7}$~km). The best-fit spectral parameters and spectra are shown in Table~1 and Figure~3, respectively. The pre-burst spectra were further explored using a BB+BB model which provided acceptable fits ($\chi^2_{\nu}$=0.942 (1267)), but with a lower column density $N_H$=(1.9$\pm$0.1)$\times$10$^{22}$ cm$^{-2}$, soft $kT$=0.47$^{+0.06}_{-0.05}$~keV, and hard $kT$=1.46$^{+0.31}_{-0.58}$~keV.  Here, we adopt PL+BB as the best-fit model since the $N_H$ derived in this case is closer to that obtained for its associated SNR Kes 73 (Kumar et al. 2010, in preparation). The spectral parameters obtained for the pre-burst quiescent emission are in good agreement with those obtained in previous studies (Morii et al. 2003; Nakagawa et al. 2009).

The 0.5--10 keV unabsorbed pre-burst and post-burst fluxes are $F_{XMM}$=4.3$^{+0.9}_{-1.2}$$\times$10$^{-11}$ ergs~cm$^{-2}$~s$^{-1}$ ($L_{XMM}$=3.7$^{+0.8}_{-1.0}$$\times$10$^{35}$ ergs~s$^{-1}$) and $F_{XRT}$=9.1$^{+1.2}_{-1.4}$$\times$10$^{-11}$ ergs~cm$^{-2}$~s$^{-1}$  
($L_{XRT}$=7.8$^{+1.0}_{-1.2}$$\times$10$^{35}$ ergs~s$^{-1}$), respectively. This represents a 2.1 times increase in the unabsorbed flux following the burst, consistent with the pulsed flux increase (2.02$\pm$0.06) reported using a Rossi X-ray Timing Explorer (\textit{RXTE}) ToO observation in the 2--11 keV band (Dib et al. 2010).

\section{Discussion}
\label{4}

Thanks to \textit{Swift},  the first burst from 1E~1841--045 and outbursts from two other AXPs (CXOU~J164710.2--455216, 1E~1541.0--5408) have been detected, enabling a detailed study of their burst activity with BAT in the hard X-ray band and of the underlying prompt persistent emission immediately after outburst with XRT in the soft X-ray band.  Such studies are vital to understanding the physics of the outburst and for testing the predictions of the magnetar model.

As an extension to the magnetar model, Thompson et al. (2002) suggest that the twisted internal $B$-field stresses the crust in turn twisting the external dipole field. When a static twist is implanted, currents flow into the magnetosphere. As the twist angle grows, electrons provide an increasing optical depth to resonant cyclotron scattering building up a flatter photon power-law component. Meanwhile, returning currents provide an extra heating of the star surface increasing the X-ray flux. Hence, a correlation between X-ray flux and spectral hardness is expected. Bursts arise from the sudden small-scale surface reconfiguration owing to a magnetospheric twist and the activity increases with increasing twist angle. Beloborodov (2009) suggests that the twisted magnetosphere gradually untwists by producing radiation where the thermal component is expected to survive the time-scale required to dissipate the twist energy, while the non-thermal component is short-lived since the resonant scattering is no longer possible when the current-carrying bundle becomes too small.

The burst observed from 1E~1841--045 is short (32~ms), symmetric, and well-fit by a PL model in the 15--100 keV range. These characteristics fit the description of Type A AXP bursts, similar to those seen in SGRs, which are short, symmetric, and uncorrelated with pulse phase as opposed to type B bursts, seen exclusively in AXPs, with long extended tails (lasting tens to hundreds seconds), thermal spectra, and occurring at pulse maximum (Woods et al. 2005). Furthermore, Type A and B bursts are believed to be produced by different mechanisms with the former interpreted as due to a reconnection in the upper magnetosphere (Lyutikov 2002) while the latter predominantly due to a rearrangement of the $B$-field lines anchored to the surface after a crustal fracture (Thompson \& Duncan 1995). It is also notable that 1E~1841--045, showing hard X-ray emission up to 150 keV ($\Gamma$=1.32$\pm$0.11; Kuiper et al. 2004) and interpreted as originating from the magnetosphere, exhibited a soft burst spectrum ($\Gamma$=3.2$^{+1.8}_{-1.0}$; 15--100 keV) possibly due to a renewed magnetospheric activity in the external $B$-field.

We studied the 0.5--10 keV pre- and post-burst persistent emission of 1E~1841--045 using \textit{XMM-Newton} and \textit{Swift}-XRT data, respectively. The \textit{XMM-Newton} spectra were described by a BB-component ($kT$=0.45$\pm$0.03~keV, $R_{BB}$=5.0$^{+0.5}_{-0.7}$~km) possibly originating from a hot spot on the NS surface, plus a PL-component ($\Gamma$=1.9$\pm$0.2) likely associated with the magnetosphere. Our XRT spectral analysis reveals that the source spectrum softened during the burst with the persistent spectra well fitted by a single PL model ($\Gamma$=2.6$\pm$0.2) accounting for the total flux, which increased by 2.1 times compared to its pre-burst value. By including a BB-component to the PL model (though not required statistically; Table~1), we find that the PL flux increased by 35$\%$ with an increase in $\Gamma$, while the BB flux decreased by 22$\%$ with a slight increase in temperature ($kT$=0.55$^{+0.23}_{-0.20}$~keV, $R_{BB}$=1.6$^{+2.0}_{-1.6}$~km) with respect to its pre-burst values.  If the blackbody emission originates from a hot spot, we find that it has become slightly hotter and smaller post-burst, possibly associated with the burst activity following a small-scale rearrangement of the $B$-field. But, overall the spectrum softened and the total flux increased.

Next, we compare our results with those seen in the two other \textit{Swift} observed AXPs: CXOU~J164710.2--455216 and 1E~1541.0--5408. The CXOU~J164710.2--455216 burst appears similar to that seen in 1E~1841--045 in terms of the light curve (symmetric), duration ($\sim$20~ms), and energy ($\sim$10$^{37}$ ergs). The XRT observations taken $\sim$13 hrs past the burst showed a dominant PL-component for $\sim$10 days with the BB-component dominating (80$\%$--90$\%$) the total flux in the observations taken a month later (Israel et al. 2007). \textit{XMM-Newton} observations taken 4.3 days prior to and 1.5 days post this outburst suggest spectral hardening accompanied by 100 times luminosity increase, interpreted as due to a plastic deformation of the NS's crust which induced a slight twist in the external $B$-field causing the X-ray burst (Muno et al. 2007).  However, a reconnection activity in the 1E~1841--045 magnetosphere might have caused the X-ray burst. On the other hand,  the bursts from 1E~1541.0--5408 were characterized by long extended tails in some of its 2008 bursts and the XRT spectra (spanning 100~s since the BAT trigger until 3 weeks post-burst)  were described by either a PL or a BB model, with the initial hard ($\Gamma$$\sim$2 or $kT$$\sim$1.4~keV) outburst spectrum steepening ($\Gamma$$\sim$4 or $kT$$\sim$0.8~keV) within one day from the BAT trigger (Israel et al. 2010). When fitted with a BB+PL model, the spectrum taken 100~s after the BAT trigger was totally dominated by a PL-component, while in the observations taken 0.05 and 0.2 days after the trigger, the BB-component becomes dominant and one day later, the PL-component becomes undetectable (Israel et al. 2010). Similarly, the 1E~1841--045 post-burst emission, taken 0.5~hr and 1.9~hr since the BAT trigger, is also dominated by the PL-component even though the timescales are different. Unfortunately, XRT stopped observing 1E~1841--045 within 3~hrs of the BAT trigger and hence, we cannot make a judgement about the evolution of its spectrum in comparison with the other two AXPs. However we note that while both CXOU~J164710.2--455216 and 1E~1541.0--5408 had shown a flux-hardness correlation associated with the burst, we see a flux-hardness anti-correlation in 1E~1841--045 (or steepening of the spectrum immediately post-burst), a result that is contradictory to the twisted magnetosphere predictions.  However, in a globally twisted magnetosphere, the X-ray spectrum can soften following the burst as the magnetosphere becomes more transparent to cyclotron scattering (Thompson et al. 2002).

The \textit{Swift}-BAT burst spectrum of 1E~1841--045 is further intriguing in that it showed emission line features at energies 26.7$^{+1.6}_{-1.4}$~keV, 39.6$^{+2.8}_{-2.2}$~keV and 59.8$^{+3.1}_{-3.5}$~keV, at the $\sim$2--3 sigma level. Magnetars' spectral features (often interpreted as proton cyclotron lines from $B$-fields$\sim$10$^{14}$--10$^{15}$~G) have been reported, although not always with high statistical significance, using \textit{RXTE} from SGRs 1900+14, 1806--20, and AXPs XTE J1810--197, 1E~1048.1--5937,  4U~0142+61 (Strohmayer \& Ibrahim 2000; Ibrahim et al. 2002; Woods et al. 2005; Gavriil et al. 2002, 2010). 
The features reported here are interestingly multiples of the 13--14 keV line reported with $RXTE$ in 3 other AXPs; and so, if interpreted as the second, third, and fourth harmonics of a proton cyclotron line at $\sim$13--14 keV, they would yield $B$=(3.5--3.7)$\times$10$^{15}$~G, which is close to the AXP's dipole $B$-field (7.1$\times$10$^{14}$~G). However, the detection of harmonics would argue against a proton cyclotron origin since higher harmonics should be suppressed for proton resonances (S. Zane, private communication). On the other hand, electron cyclotron lines from magnetar-strength $B$-fields would fall in the MeV energy range.

We have further investigated whether any of these lines could be instrumental background and consulted with the BAT team (C. Markwardt, private communication). The BAT CdZnTe detectors have escape lines and instrumental K-edges at 27 keV (Cd), 32 keV (Te), and $^{241}$Am lines at 59.5~keV\footnote{http://heasarc.nasa.gov/docs/swift/analysis/bat$\_$digest.html.}. While the mask-weighting technique and systematic error correction should take care of these lines, they can be still present though will be only a percentage of the source flux; and for a short burst of 32~ms, background line features will be negligible.  We however note that the background subtraction that comes from mask-weighting depends on the assumption of Gaussian statistics, which likely breaks down at low count levels. Unfortunately, the low statistics limit is a realm which the \textit{BAT} team has not explored for understanding statistics of the mask-weighting technique (C. Markwardt, private communication). We conclude that the origin of these spectral features is unclear and a confirmation of their presence is needed with other instruments.

\section{Conclusion}
\label{5}

We have presented a detailed analysis of the \textit{Swift} observations of the first burst detected from the AXP 1E~1841--045. The 15--100 keV time-averaged burst spectrum and the 0.5--10 keV persistent spectra obtained with \textit{Swift} were described by a single PL model, both showing a softer spectrum ($\Gamma$=2.6--3.2) than its pre-burst spectrum obtained with \textit{XMM-Newton} ($\Gamma$=1.9). We conclude that the source has softened post-burst as seen from the XRT observations taken within $\leq$3 hrs since the BAT trigger, with a 2.1 times flux increase compared to its pre-burst value in the 0.5--10 keV range. We discussed our findings in the light of the magnetar model predictions and in comparison with other magnetar bursts. We also reported on emission features observed in the \textit{Swift}-BAT burst spectrum. Observations with other instruments during active burst phases are warranted to confirm the existence of such lines and to understand their nature.

\acknowledgments

This research made use of NASA's ADS and HEASARC maintained at GSFC. SSH acknowledges support by NSERC and the Canada Research Chairs  program. We thank Silvia Zane for comments on the manuscript, C. Markwardt for discussions on the BAT spectral features, and K. Arnaud  for discussions on the Monte Carlo simulation.


\begin{thebibliography}{}

\bibitem{Barthelmy:2005}
Barthelmy, S. D., et al. 2005, Space Science Reviews, 120, 143
\bibitem{Beardmore:2010}
Beardmore, A. P., et al. 2010, GCN, 10722
\bibitem{Belo:2009}
Beloborodov, A. M. 2009, \apj, 703, 1044
\bibitem{Burrows:2005}
Burrows, D. N., et al. 2005, Space Sci. Rev., 120, 165
\bibitem{Dib:2008}
Dib, R., Kaspi, V. M., \& Gavriil, F. P. 2008, \apj, 673, 1044
\bibitem{Dib:2010}
Dib, R., Kaspi, V. M., \& Gavriil, F. P. 2010, ATel, 2602
\bibitem{Gavriil:2002}
Gavriil, F. P., Kaspi, V. M., \& Woods, P. M. 2002, Nature, 419, 142
\bibitem{Gavriil:2010}
Gavriil, F. P., Dib, R., \& Kaspi, V. M. 2010, \apj, submitted (arXiv:0905.1256)
\bibitem{gehrels2004}
Gehrels, N., et al. 2004, \apj, 611, 1005 
\bibitem{Ibrahim:2002}
Ibrahim, A. I., Safi-Harb, S., Swank, J. H., Parke, W., Zane, S., \& Turolla, R. 2002, \apjl, 574, 51
\bibitem{Israel:2007}
Israel, G. L., Campana, S., Dall'Osso, S., Muno, M. P., Cummings, J., Perna, R., \& Stella, L. 2007, \apj, 664, 448
\bibitem{Israel:2010}
Israel, G. L., et al. 2010, MNRAS, 408, 1387
\bibitem{Kaspi2003}
Kaspi, V. M., Gavriil, F. P., Woods, P. M., Jensen, J. B., Roberts, M. S. E., \& Chakrabarty, D. 2003, \apjl, 588, 93
\bibitem{Kuiper:2004}
Kuiper, L., Hermsen, W., Mendez M., 2004, \apj, 613, 1173 
\bibitem{Lyutikov:2002}
Lyutikov, M. 2002, \apjl, 580, 65
\bibitem{Mereghetti:2008}
Mereghetti, S. 2008, A\&AR, 15, 225
\bibitem{Morri:2003}
Morii, M., Sato, R., Kataoka, J., \& Kawai, N. 2003, PASJ,  55, L45
\bibitem{Muno:2007}
Muno, M. P., Gaensler, B. M., Clark, J. S., de Grijs, R., Pooley, D., Stevens, I. R., Portegies Zwart, S. F. 2007, MNRAS, 378, L44
\bibitem{Nakagawa:2009}
Nakagawa, Y. E., Yoshida, A., Yamaoka, K., \&  Shibazaki, N. 2009, PASJ, 61, 109
\bibitem{Stroh:2000}
Strohmayer, T. E., \& Ibrahim, A. I. 2000, \apjl, 537, 111
\bibitem{Struder:2001}
Struder, L., et al. 2001, A\&A, 365, L18
\bibitem{TianLeahy:2008}
Tian, W. W., \& Leahy, D. A. 2008, \apj, 677, 292
\bibitem{ThompsonDuncan:1995}
Thompson, C., \& Duncan, R. C. 1995, MNRAS, 275, 255
\bibitem{Thompson:2002}
Thompson, C., Kyutikov, M., \& Kulkarni, S. R. 2002, \apj, 574, 332
\bibitem{Turner:2001}
Turner, M. J. L., et al. 2001, A\&A, 365, L27
\bibitem{Vasisht:1997}
Vasisht, G., \& Gotthelf, E. V. 1997, \apjl, 486, 129 
\bibitem{Woods:2005}
Woods P. M., et al. 2005, \apj, 629, 985
\bibitem{Zhukaspi:2010}
Zhu, W., \& Kaspi, V. M. 2010, \apj, 719, 351

\end{thebibliography}
\end{document}